\documentclass[pre,preprint,superscriptaddress]{revtex4}

\usepackage{graphicx}
\usepackage{color}
\usepackage{amsmath,amssymb}
\usepackage{mathrsfs}

\begin{document}

\title{Underdamped scaled Brownian motion: (non-)existence of the overdamped
limit in anomalous diffusion}

\author{Anna S. Bodrova}
\affiliation{Institut f\"ur Physik, Humboldt-Universit\"at zu Berlin,
Newtonstrasse
15, 12489 Berlin, Germany}
\affiliation{Faculty of Physics, M.V.Lomonosov Moscow State University, Moscow,
119991, Russia}
\author{Aleksei V. Chechkin}
\affiliation{Akhiezer Institute for Theoretical Physics, Kharkov Institute of
Physics and Technology, Kharkov 61108, Ukraine}
\affiliation{Institute of Physics and Astronomy, University of Potsdam, 14476
Potsdam, Germany}
\affiliation{Department of Physics \& Astronomy, University of Padova, 35122
Padova, Italy}
\author{Andrey G. Cherstvy}
\affiliation{Institute of Physics and Astronomy, University of Potsdam, 14476
Potsdam, Germany}
\author{Hadiseh Safdari}
\affiliation{Department of Physics, Shahid Beheshti University, G.C., Evin,
Tehran 19839, Iran}
\affiliation{Institute of Physics and Astronomy, University of Potsdam, 14476
Potsdam, Germany}
\author{Igor M. Sokolov}
\affiliation{Institut f\"ur Physik, Humboldt-Universit\"at zu Berlin,
Newtonstrasse 15, 12489 Berlin, Germany}
\author{Ralf Metzler}
\thanks{Correspondence to rmetzler@uni-potsdam.de}
\affiliation{Institute of Physics and Astronomy, University of Potsdam, 14476
Potsdam, Germany}

\begin{abstract}
It is quite generally assumed that the overdamped Langevin equation provides a
quantitative description of the dynamics of a classical Brownian particle in the
long time limit. We establish and investigate a paradigm anomalous diffusion
process governed by an underdamped Langevin equation with an explicit time
dependence of the system temperature and thus the diffusion and damping
coefficients. We show that for this underdamped scaled Brownian motion
(UDSBM) the overdamped limit fails to describe the long time behaviour
of the system and may practically even not exist at all for a certain range of
the parameter values. Thus persistent inertial effects play a non-negligible role
even at significantly long times. From this study a general questions on the
applicability of the overdamped limit to describe the long time motion of
an anomalously diffusing particle arises, with profound consequences for
the relevance of overdamped anomalous diffusion models. We elucidate our
results in view of analytical and simulations results for the anomalous
diffusion of particles in free cooling granular gases.
\end{abstract}

\maketitle

The mean squared displacement (MSD) of a Brownian particle at sufficiently long
times follows the linear time dependence $\langle x^2(t)\rangle\simeq K_1t$, as
predicted by the second Fick's law \cite{fick} and physically explained by
Einstein \cite{einstein} and Smoluchowski \cite{smoluchowski}. However, already
in 1926 Richardson reported the distinct non-Fickian behaviour of tracer particles
in atmospheric turbulence \cite{richardson}. Today, such \emph{anomalous
diffusion\/} is typically associated with the power-law form
\begin{equation}
\left< x^2(t) \right> \simeq t^\alpha
\label{x2-AD}
\end{equation}
of the MSD, where subdiffusion corresponds to values of the anomalous diffusion
exponent $\alpha$ in the range $0<\alpha<1$ and superdiffusion to $\alpha>1$
\cite{Bouchaudrev,MetzlerREVIEW,sokolovSM,MetzlerREVIEWpccp14}. Classical examples
for subdiffusion include the charge carrier motion in amorphous semiconductors
\cite{scher}, the spreading of tracer chemicals in subsurface aquifers \cite{brian}
or in convection rolls \cite{tabeling}, as well as the motion of a tracer particle
in a single file of interacting particles \cite{singlefile}. Superdiffusion is
known from tracer motion in turbulent flows \cite{richardson} and weakly chaotic
systems \cite{swinney}, or for randomly searching, actively moving creatures such
as microorganisms and bacteria \cite{beer15}, albatrosses \cite{sims2012}, or
humans \cite{brockmann}.

Modern microscopic techniques, in particular, superresolution microscopy, have led
to the discovery of a multitude of anomalous diffusion processes in living
biological cells and complex fluids \cite{MetzlerPT,hoefling,MetzlerREVIEWpccp14,
sokolovsub}. Thus subdiffusion was observed in live cells for RNA molecules
\cite{golding}, chromosomal telomeres \cite{garini}, or submicron lipid \cite{lene}
and insulin granules \cite{tabei}. Even small proteins such as GFP were
demonstrated to subdiffuse \cite{gratton}. In artificially crowded systems,
subdiffusion is also routinely observed \cite{weiss1,pan,lene1,crowd2}.
Superdiffusion of injected as well as endogenous submicron particles, due to
active processes such as molecular motor driven transport was reported in the
cellular context \cite{elbaum,wilhelm,christine}. Following the progress of
supercomputing capabilities,
subdiffusion was also reported for complex molecular systems such as relative
diffusion in single proteins \cite{smith}, in pure \cite{kneller,jae_membrane}
and crowded \cite{faraday,prx} lipid bilayer membranes \cite{bba}.

Apart from the power-law anomalous diffusion (\ref{x2-AD}) ultraslow processes
with a logarithmic time dependence
\begin{equation}
\left< x^2(t) \right> \simeq \log t
\label{x2log}
\end{equation}
of the MSD
exist in a variety of systems \cite{MetzlerREVIEWpccp14}. Such logarithmic time
dependencies occur in Sinai diffusion in quenched random energy landscapes
\cite{sinai,sinaijpa}, periodically iterated maps \cite{maps}, colloidal hard
sphere systems at the liquid-glass transition \cite{sperl}, random walks on bundled
structures \cite{bundled}, or in single file diffusion with power-law trapping
time distributions for individual particles \cite{lloyd}. A particular system in
which ultraslow diffusion occurs are granular gases in the homogeneous cooling
stage, in which each particle-particle collision reduces the kinetic energy of
the two particles by a constant factor, the so called restitution coefficient
\cite{brilbook}.

The nature of
anomalous diffusion of the forms (\ref{x2-AD}) or (\ref{x2log}) is non-universal
and may originate from numerous physical processes. Power-law anomalous diffusion,
for instance, emerges for continuous time random walk processes with scale-free
distributions of waiting times or jump lengths \cite{scher,klablushle}, generalised
Langevin equations of fractional Brownian motion with power-law correlated,
Gaussian noise input \cite{fle}, or diffusion processes with deterministic
\cite{hdp} or random \cite{lapeyre} position dependence of the diffusivity.
Ultraslow diffusion can be described in terms of continuous time random walks with
super heavy-tailed waiting times \cite{havlin,sinaijpa} or heterogeneous diffusion
processes with exponential space dependence of the diffusivity \cite{HDP-PCCP}.

The motion of a particle of mass $m$ in a thermal bath is typically described by
a Langevin equation \cite{langevin,vankampen}. While the short time motion of
this particle is ballistic, once collision events become relevant, a crossover
to normal Brownian motion with MSD (\ref{x2-AD}) and $\alpha=1$ occurs. The
corresponding crossover time scale is given by the inverse
friction coefficient. For Brownian motion at sufficiently long times it is
sufficient to use the overdamped Langevin equation without the inertia term, to
quantitatively describe the particle motion. In other words, the long
time limit of the full Langevin equation including the Newton term $m\ddot{x}(t)$
coincides with the solution of the overdamped Langevin equation \cite{vankampen,
risken}. 

Here we study a simple anomalous diffusion process based on the full Langevin
equation with inertial term and a time dependent diffusion coefficient. For
this underdamped scaled Brownian motion (UDSBM) we demonstrate that the long
time limit may be distinctly disparate from the analogous overdamped process
due to extremely persistent inertial effects, that dominate the particle motion
on intermediate-asymptotic time scales.
This a priori surprising finding breaks with a commonly accepted dogma for
stochastic processes and demonstrates that the correct mathematical description
for particles with a mass in the long time limit for
anomalous diffusion processes may be a delicate issue, that requires special care.
Our findings are based on analytical calculations and confirmed by extensive
stochastic simulations. Comparison to event driven simulations of granular
gases confirm the results of our UDSBM model for a physical model based on first
principles.

To proceed, we first provide a concise summary of the properties of the regular
underdamped Langevin equation for Brownian motion and its overdamped limit. The
following Section then briefly introduces the overdamped Langevin description for
scaled Brownian motion (SBM) corresponding to the UDSBM process without the inertia
term. The subsequent section then introduces the full Langevin equation for UDSBM
including the mass term. We unravel the ensemble and time averaged characteristics
of this UDSBM process analytically and show the agreement with stochastic
simulations. Both cases of power-law anomalous diffusion (\ref{x2-AD}) as well as
ultraslow diffusion (\ref{x2log}) are considered. In particular, we also present
a comparison of the UDSBM process with event driven simulations of a cooling
granular gas. Mathematical details of the derivations are presented in the
Methods section.

\section*{LANGEVIN EQUATION WITH CONSTANT COEFFICIENTS}

In this section we briefly recall the basic properties of the stochastic
description of Brownian motion, in particular, the transitions from the
under- to the overdamped regimes. We consider both the more traditional
ensemble averages of moments and the corresponding time averages,
important for the analysis of time series obtained from particle tracking
experiment and simulations \cite{MetzlerREVIEWpccp14,MetzlerPT}.

\subsection*{Overdamped Langevin equation}

Let us start with the overdamped Langevin equation with the constant diffusion
coefficient $D_0$ \cite{risken,vankampen},
\begin{equation}
\frac{dx(t)}{dt}=v(t)=\sqrt{2D_0}\times\zeta(t),
\label{usbm000}
\end{equation}
fuelled by the Gaussian $\zeta(t)$ with $\delta$-correlation
\begin{equation}
\left\langle\zeta(t_1)\zeta(t_2)\right\rangle=\delta\left(t_2-t_1\right)
\end{equation}
and zero mean $\langle\zeta(t)\rangle=0$. The corresponding MSD has the linear
time dependence
\begin{equation}
\left\langle x^2(t)\right\rangle=2D_0t
\label{MSDnormal}
\end{equation}
expected for overdamped Brownian motion of a test particle in a thermal bath. The
noise strength is given by the diffusion constant $D_0$.

In the single particle tracking experiments and massive computer simulations often
only few but long traces are available for the analysis. In this case one typically
analyses the particle motion encoded in the time series $x(t)$ via the time averaged
MSD \cite{MetzlerREVIEWpccp14,MetzlerPT}
\begin{equation}
\overline{\delta^2(\Delta)}=\frac{1}{t-\Delta}\int_0^{t-\Delta}\Big[x(t'+\Delta)
-x(t')\Big]^2dt'.
\label{deltadef}
\end{equation}
Here $\Delta$ is the lag time and $t$ denotes the total length of the trajectory
(measurement time). An additional average over $N$ time traces $x_i(t)$
\begin{equation}
\left<\overline{\delta^2(\Delta)}\right>=\frac{1}{N}\sum_{i=1}^{N}\overline{
\delta^2_i(\Delta)}
\label{tadef}
\end{equation}
then produces a smooth variation of the time averaged MSD with the lag time. For
Brownian motion we observe the equality
$\left<[x(t'+\Delta)-x(t')]^2\right>\sim\langle\delta x^2\rangle\times\Delta/
\tau$, where $\langle\delta x^2\rangle$ is the variance of
the underlying jump length distribution, and $\tau$ is the typical time for a
single jump \cite{MetzlerREVIEWpccp14,MetzlerPT}. We therefore obtain the
equality
\begin{equation}
\left<\overline{\delta^2(\Delta)}\right>=\left\langle x^2(\Delta)\right\rangle=2
D_0 \Delta,
\end{equation}
so that the system is ergodic in the Boltzmann-Khinchin sense, that is, time and
ensemble averages coincide. In particular, we see that the time averaged MSD
$\left<\overline{\delta^2(\Delta)}\right>$ is independent of the observation time
$t$, reflecting the stationarity of the process.

\subsection*{Underdamped Langevin equation}

Now consider the underdamped Langevin equation with inertial term \cite{risken,
vankampen},
\begin{equation}
\frac{d^2x(t)}{dt^2}+\gamma_0\frac{dx(t)}{dt}=\sqrt{2D_0}\times\gamma_0\zeta(t).
\label{usbm0}
\end{equation}
The constant damping coefficient $\gamma_0$ and the diffusion coefficient $D_0$
are connected via the Einstein-Smoluchowski-Sutherland fluctuation dissipation
relation
\begin{equation}
D_0=\frac{T_0}{m\gamma_0},
\end{equation}
where we use the convention to set the Boltzmann constant $k_B$ to unity. The
two point velocity correlation function encoded by the underdamped Langevin
equation (\ref{usbm0}) decays exponentially in the time difference,
\begin{equation}
\left\langle v(t_1)v(t_2)\right\rangle=\frac{T_0}{m}\exp\Big(-\gamma_0\left|t_2-
t_1\right|\Big).
\end{equation}
The associated characteristic time is defined by the inverse of the friction
coefficient, $1/\gamma_0$. The MSD follows from the velocity correlation function
via
\begin{eqnarray}
\nonumber
\left\langle x^2(t)\right\rangle&=&2\int_0^t dt_1 \int_{0}^{t-t_1} d\Delta t
\langle v(t_1)v(t_1+\Delta t)\rangle\\
&=&2D_0t +\frac{2D_0}{\gamma_0}\left(e^{-\gamma_0 t}-1\right).
\label{msd0}
\end{eqnarray}
At short times $t\ll 1/\gamma_0$ the MSD scales ballistically, $\left\langle x^2
(t)\right\rangle\simeq D_0\gamma_0 t^2$ while at long times $t\gg 1/\gamma_0$ the
MSD is given by the linear time dependence (\ref{MSDnormal}) of the overdamped
Langevin equation. Thus the inertial effects indeed cancel out rapidly and are
important only at times smaller than or comparable to the characteristic time
scale $1/\gamma_0$.

For the underdamped Langevin equation the time averaged MSD is calculated using
Eqs.~(\ref{deltadef}) and (\ref{tadef}). It has the same time dependence as the
ensemble averaged MSD, namely,
\begin{equation}
\left<\overline{\delta^2(\Delta)}\right>=\left\langle x^2(\Delta)\right\rangle=
2D_0\Delta+\frac{2D_0}{\gamma_0}\Big(e^{-\gamma_0 \Delta}-1\Big).
\label{ddd0}
\end{equation}
In addition to this ergodic behaviour, we have thus corroborated that the
dynamic encoded in the overdamped Langevin equation (\ref{usbm000}) exactly
equals the long time limit of the underdamped Langevin equation (\ref{usbm0}).

\section*{SCALED BROWNIAN MOTION}

Scaled Brownian motion (SBM) designates an anomalous diffusion process based on
an overdamped Langevin equation fuelled by white Gaussian noise, see below. SBM
involves a power law time dependent diffusion coefficient $D(t)\simeq t^{\alpha
-1}$ \cite{LimSBM,MetzlerSBM,SokolovSBM,hadisehEB,hadiseh}, stemming from a time
dependence of the system temperature, see below. SBM is a quite simple
process, as it is Markovian. Concurrently, it is strongly non-stationary. For
this reason the process stays time dependent even in a confining external
potential and is weakly non-ergodic as well as ageing in the sense defined below
\cite{MetzlerSBM,SokolovSBM,hadisehEB,hadiseh}.

SBM should not be confused with fractional Langevin equation motion or fractional
Brownian motion which are non-Markovian yet Gaussian processes with stationary
increments whose
probability density in the overdamped limit coincides with that of SBM but has a
completely different physical origin \cite{peter,fle,MetzlerREVIEWpccp14}.
The underdamped Langevin equation for fractional Langevin equation motion was
analysed in Refs.~\cite{eli,jae_pre1,kursawe,lene1} and shown to exhibit
interesting effects such as oscillatory behaviour of the velocity correlations
as well as transient ageing
and non-ergodic behaviour. However, these decay rather quickly to make way for
the expected overdamped behaviour. Here we show that the behaviour of UDSBM is
significantly different from the fractional Langevin equation motion and involved
persistent inertial terms.

Before starting the discussion of SBM we note that anomalous diffusion with time
dependent diffusion coefficient $D(t)\simeq t^{\alpha-1}$ occurs, for instance,
in the famed Batchelor model for turbulent diffusion \cite{batchelor}. SBM was
used to model the water diffusion in brain measured by magnetic resonance imaging
\cite{novikov}, the mobility of proteins in cell membranes \cite{biophys1996}, or
the motion of molecules in porous environments \cite{Sen}. As effective
subdiffusion model it was also used to describe biological systems \cite{saxton,
schwille,weiss}. Physically time
dependent diffusion coefficients arise naturally in systems with a time dependent
temperature such as melting snow \cite{molini,snow} or free cooling granular gases,
in which the temperature is given by the kinetic energy, which dissipates
progressively into internal degrees of freedom of the gas particles
\cite{brilbook,ggg1,ggg2}.

\subsection*{Scaled Brownian motion with $\alpha>0$}

The overdamped SBM Langevin equation with time dependent diffusion coefficient
$D(t)\simeq t^{\alpha-1}$ and $\alpha>0$ is typically used as the definition of
SBM \cite{LimSBM,MetzlerSBM,SokolovSBM,hadisehEB,hadiseh}
\begin{equation}
\frac{dx(t)}{dt}=\sqrt{2D(t)}\times\zeta(t).
\label{lang_sbm}
\end{equation} 
Here we consider the time dependent diffusion coefficient in the more general form
\begin{equation}
\label{diffcoef}
D(t)=D_0\left(1+t/\tau_0\right)^{\alpha-1},
\end{equation} 
which avoids a singular behaviour at $t=0$, and $\tau_0$ represents a
characteristic time for the mobility variation. For this choice $D_0=D(0)$ is the
initial diffusion coefficient. The specific form (\ref{diffcoef}) of
$D(t)$ is primarily motivated by the corresponding expression derived in the theory
of cooling
granular gases \cite{annagg}. In addition Eq.~(\ref{diffcoef}) represents a
simple smooth function allowing us to reproduce all three regimes in the evolution
of the MSD we are interested in in what follows, namely, ballistic, normal, and
anomalous.

Given definition (\ref{diffcoef}) the mean squared displacement follows in the form
\begin{eqnarray}
\left\langle x^2(t)\right\rangle=2\int_0^{t}D(t')dt'=\frac{2D_0\tau_0}{\alpha}
\left(\left(1+\frac{t}{\tau_0}\right)^{\alpha}-1\right).
\label{msdsbm}
\end{eqnarray}
Thus the MSD grows linearly, $\left\langle x^2(t)\right\rangle\sim2D_0t$ at short
times $t\ll\tau_0$. At long times $t\gg\tau_0$ it scales according to
Eq.~(\ref{x2-AD}) and thus covers both sub- and superdiffusive processes
\cite{LimSBM,MetzlerSBM,SokolovSBM,hadisehEB,hadiseh}.

The full expression for the time averaged MSD is given by Eq.~(\ref{delta0}) in
the Methods section. At short times $\Delta\ll t\ll\tau_0$
the diffusion coefficient is almost unchanged, $D(t)\approx D_0$ and normal ergodic
behaviour is observed,
$\left\langle\overline{\delta_0^2(\Delta)}\right\rangle\simeq\left\langle x^2(
\Delta)\right\rangle\simeq2D_0\Delta$. At longer lag times $\tau_0\ll\Delta\ll t$
we get that
\begin{equation}
\left\langle\overline{\delta_0^2(\Delta)}\right\rangle\simeq\frac{2D_0\Delta}{
\alpha \left(t/\tau_0\right)^{1-\alpha}}.
\label{deltaUDSBMover}
\end{equation}
Thus the MSD and the time averaged MSD exhibit a fundamentally different (lag)
time dependence, a weak breaking of ergodicity. In contrast to the Langevin
equation with constant coefficients the time averaged MSD now also depends on
the measurement time $t$, a phenomenon called ageing \cite{MetzlerREVIEWpccp14}.

\subsection*{Ultraslow SBM with $\alpha=0$}

Ultraslow SBM corresponds to the limiting case $\alpha=0$ for the diffusion
coefficient (Eq.~\ref{diffcoef}) \cite{ultraslow},
\begin{equation}
\label{diffcoef0}
D(t)=D_0\left(1+t/\tau_0\right)^{-1}.
\end{equation} 
In this case the MSD has the logarithmic time dependence 
\begin{equation}
\left\langle x^2(t)\right\rangle=2D_0\tau_0\log\left(1+\frac{t}{\tau_0}\right)\,.
\label{msdsbm0}
\end{equation}
At long times the MSD $\left\langle x^2(t)\right\rangle$ converges to
Eq.~(\ref{x2log}). The full expression for the time averaged MSD is given by
Eq.~(\ref{delta0ultraslow}) in Methods. For $\tau_0\ll\Delta\ll
t$ the time averaged MSD has the following mixed power-law-logarithmic scaling
\cite{ultraslow}
\begin{equation}
\label{dusbm0}
\left\langle\overline{\delta_0^2(\Delta)}\right\rangle\simeq 2D_0\tau_0\frac{
\Delta}{t}\log\left(\frac{t}{\Delta}\right),
\end{equation}
which again features an ageing behaviour \cite{hadisehEB,hadiseh}.
At short times $\Delta\ll\tau_0$, $t\ll\tau_0$ normal diffusion is observed,
$\left\langle\overline{\delta_0^2(\Delta)}\right\rangle\simeq\left\langle x^2(
\Delta)\right\rangle\simeq2D_0\Delta$.

\section*{RESULTS}

\section*{UNDERDAMPED SCALED BROWNIAN MOTION}

Let us now turn to the UDSBM case and consider the underdamped version of the
Langevin equation (\ref{lang_sbm}) with time dependent diffusion and damping
coefficients, $D(t)$ and $\gamma(t)$, respectively,
\begin{equation}\label{Lang}
\frac{d^2x(t)}{dt^2}+\gamma(t)\frac{dx(t)}{dt}=\sqrt{2D(t)}\gamma(t)\zeta(t).
\end{equation}
In that sense it is a straightforward extension of the Brownian Langevin
equation (\ref{usbm0}) with additional multiplicative coefficients.
We assume that the particle moves in a bath with temperature $T(t)$ with power
law time dependence
\begin{equation} 
\label{haff}
T(t)=T_0\left(1+t/\tau_0\right)^{2\alpha-2},
\end{equation}
where $\alpha\ge0$ and the value $T_0=T(0)$ is the initial temperature. The time
scale $\tau_0$ corresponds to the characteristic time of the temperature decay.
Larger $\tau_0$ values imply a slower temperature decrease. In the limit $\tau_0
=\infty$ the temperature of the system remains constant, which corresponds to the
case of normal diffusion. We assume that the bath is in local equilibrium, and the
time dependent damping coefficient scales as $\gamma(t)\simeq\sqrt{T(t)}$ or
\begin{equation}
\label{damp}
\gamma(t)=\gamma_0\left(1+t/\tau_0\right)^{\alpha-1}
\end{equation} 
with the initial value $\gamma_0=\gamma(0)$. Thus $1/
\gamma(t)$ defines the characteristic decay time of the velocity correlation
function, which is now also time dependent. The choice of the damping coefficient
in the form (\ref{damp}) appears natural since it is in accordance with the two
paradigmatic models. The first one corresponds to a massive Brownian particle in
a gas with continuous heating or cooling, consisting of elastically colliding
particles: in this
case the damping coefficient may be derived as a Stokes friction coefficient and
is proportional to the dynamical viscosity which in turn scales as $\sqrt{T}$
\cite{Klimontovich}. The second model corresponds to the self-diffusion in
granular gases. In that case the damping coefficient is equal to the inverse
velocity autocorrelation time, $\gamma(t)=\tau_v^{-1}(t)$, where $\tau_v(t)\simeq
T^{-1/2}$ \cite{brilbook}.

The time dependent diffusion coefficient may then be related to the damping
coefficient according to the (time local) fluctuation dissipation theorem
\cite{MetzlerSBM,anna_gran},
\begin{equation}
\label{lfdt}
D(t)=\frac{T(t)}{\gamma(t)m}.
\end{equation} 
This way we recover the diffusion coefficient (\ref{diffcoef}) introduced above
with the initial value $D_0=T_0/\left(\gamma_0 m\right)$. In the picture of the
cooling granular gas the decrease of the granular temperature due to dissipative
collisions of particles according to Eq.~(\ref{haff}) was indeed observed
\cite{brilbook}. Here the case $\alpha=0$
considered in subsection B corresponds to particles colliding with
constant restitution coefficient \cite{haff}, and $\alpha=1/6$ to granular gases
of viscoelastic particles colliding with relative velocity dependent restitution
coefficient \cite{brilbook}. The diffusion coefficient in the granular gases
decays according to Eq.~(\ref{diffcoef}) \cite{brilbook,seldiffusionbril,
seldiffusionbrey,breydiff,breydiff1,anna_gran,annaprl} and the motion of granular
particles slows down continuously while the inter-collision times become longer
on average. The underdamped Langevin equation (\ref{Lang}) is thus valid for both
the description of an underdamped Brownian particle in a bath with time dependent
temperature and for the self-diffusion in free cooling granular gases, as will
be elaborated further below. The Langevin approach is justified if the typical
temperature variation time scale $\tau_0$ is sufficiently larger than the inverse
initial damping coefficient, $\tau_0\gamma_0\gg1$. This time scale separation
allows us to introduce the local fluctuation dissipation theorem (\ref{lfdt}).
We stop to note that there is an alternative version of the Langevin equation
with time dependent temperature derived for a different system of a Brownian
particle interacting with a bath of harmonic oscillators \cite{breylang}.

Introducing the power-law time dependent diffusion coefficient (\ref{diffcoef})
and damping coefficient (\ref{damp}) into the Langevin equation (\ref{Lang}) we
obtain
\begin{equation}
\frac{d^2x(t)}{dt}+\frac{\gamma_0}{\left(1+\frac{t}{\tau_0}\right)^{1-\alpha}}
\frac{dx(t)}{dt}=\sqrt{2D_0}\times\gamma_0\left(1+\frac{t}{\tau_0}\right)^{3
\left(\alpha-1\right)/2}\zeta(t).
\label{usbm}
\end{equation}
We may expect that the first inertial term in this equation for subdiffusion
($\alpha\ll 1$) will behave as $v/t$ at long times, while the second term
scales as $v/t^{1-\alpha}$. For $\alpha>0$ at long measurement times $t$
the overdamped limit always dominates. However, as we will show there exists
a long lasting intermediate regime in which the motion of the particles may not
be described in terms of the overdamped approximation since both terms have
comparable contributions as long as $\alpha$ is sufficiently small. This means
that particularly for pronounced subdiffusion as in the viscoelastic granular
gas with $\alpha=1/6$ inertial effects play a significant role and
thus delay the crossover to the true overdamped limit. In contrast, for
superdiffusion this effect is negligible. In the limit of ultraslow underdamped
Langevin equation discussed below even for long times both inertial and frictional
terms have the same order of magnitude $\simeq v/t$, so the underdamped behaviour
practically dominates the entire evolution of the system. Such effects will be
clarified in detail when we consider the behaviour of MSD and time averaged MSD
below.

Before proceeding we note that the bivariate Fokker-Planck
equation (Klein-Kramers equation) corresponding to the Langevin equation
(\ref{usbm}) reads
\begin{equation}
\label{kk}
\frac{\partial}{\partial t}P(x,v,t)=\left[-\frac{\partial}{\partial x}v+
\frac{\partial}{\partial v}(\gamma(t)v)+\frac{\gamma(t)k_BT(t)}{m}\frac{
\partial^2}{\partial v^2}\right]P(x,v,t).
\end{equation}
Here $P(x,v,t)$ is the probability density function to find the text particle
with velocity $v$ at time $t$. While this equation could be solved for $P(x,v,t)$
after dual Fourier transformation in $x$ and $v$ as well as Laplace transformation
with respect to time $t$, our strategy here is based on the Langevin equation
formulation of UDSBM, as the latter allows us to immediately obtain the two-point
correlations to calculate the time averaged MSD. We also note that from the
formulation (\ref{kk}) we could read off the formal relation $D(t)=T(t)/[
m\gamma(t)]\simeq t^{\alpha-1}$ between the time-dependent diffusion coefficient
and the time-dependent temperature and friction coefficients, corresponding to
the above local fluctuation dissipation relation (\ref{lfdt}). However, we stress
again that UDSBM is an intrinsically non-stationary process off thermal
equilibrium \cite{MetzlerSBM}.

\subsection*{Underdamped scaled Brownian motion with $\alpha>0$}

\begin{figure}
\centerline{\includegraphics[width=12cm]{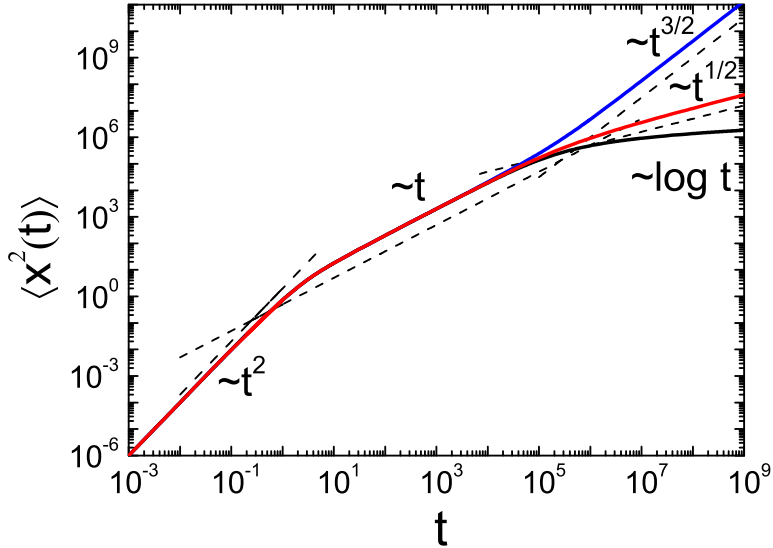}}
\caption{MSD $\left\langle x^2(t)\right\rangle$ according to Eq.~(\ref{msd}) for
$\alpha>0$ and Eq.~(\ref{R2ultraslow}) for $\alpha=0$ for the parameters $\tau_0
=100 000$, $\gamma_0=1$ with $\alpha=3/2$ (blue line), $\alpha=1/2$ (red line),
and $\alpha=0$ (black line). At short times $t\ll1/\gamma_0$ the MSD scales
ballistically, $\left\langle x^2(t)\right\rangle\simeq t^2$, at intermediate
times $1/\gamma_0\ll t\ll \tau_0$ a linear scaling $\left\langle x^2(t)\right
\rangle\sim t$ is observed, while at long times $t\gg1/\gamma_0$ the asymptotic
regime $\left\langle x^2(t)\right\rangle\sim t^{\alpha}$ is reached for $\alpha>0$,
in the case $\alpha=0$ we observe $\left\langle x^2(t)\right\rangle\simeq\log t$.}
\label{GRsupsub}
\end{figure}

We first concentrate on the details of the case $\alpha>0$. Both MSD and time
averaged MSD may be derived from the velocity correlation function, which has
the following form
\begin{eqnarray}
\left\langle v(t_1)v(t_2)\right\rangle=\frac{T_0}{m}\left(1+\frac{t_1}{\tau_0}
\right)^{2\alpha-2}\exp\left(\frac{\tau_0\gamma_0}{\alpha}\left[\left(1+\frac{
t_1}{\tau_0}\right)^{\alpha}-\left(1+\frac{t_2}{\tau_0}\right)^{\alpha}\right]
\right).
\label{velcor}
\end{eqnarray} 
The full expression for the MSD then reads
\begin{eqnarray}
\left\langle x^2(t)\right\rangle=2D_0\left[\frac{\tau_0}{\alpha}\left(\left(
1+\frac{t}{\tau_0}\right)^{\alpha}-1\right)+\frac{1}{\gamma_0}\left(\exp\left(
-\frac{\tau_0\gamma_0}{\alpha}\left[\left(1+\frac{t}{\tau_0}\right)^{\alpha}-1
\right]\right)-1\right)\right],
\label{msd}
\end{eqnarray}
which is valid as long as $\tau_0\gamma_0\gg1$, which in turn is essential for
the validity of our Langevin equation approach.
At short times corresponding to $t\ll\tau_0$ when the temperature has not changed
significantly the MSD scales according to Eq.~(\ref{msd0}). At short times $t\ll
1/\gamma_0$ compared to the scale set by the damping coefficient the MSD has the
ballistic time dependence $\left\langle x^2(t)\right\rangle\sim(T_0/m)t^2$,
which cannot be observed for the overdamped version, SBM. At intermediate times
$1/\gamma_0\ll t\ll\tau_0$ the MSD scales according to the normal diffusion law
$\left\langle x^2(t)\right\rangle\simeq 2D_0t$. At long times $t\gg\tau_0$ the
MSD follows the power-law scaling for overdamped SBM, $\left\langle x^2(t)\right
\rangle\sim2D_0\tau_0^{1-\alpha}t^{\alpha}/\alpha$. All evolution regimes are
depicted in Fig.~\ref{GRsupsub} for $\alpha=3/2$ (blue line) and $\alpha=1/2$
(red line). The ultraslow case $\alpha=0$, shown with the black line, is
considered below. It may be seen that at times $t\ll\tau_0$ the behaviour
of the MSD is independent of $\alpha$ while the $\alpha$ dependence becomes
apparent at long times.

For the derivation of the time averaged MSD we follow the same approach as
described in \cite{annagg}. It may be written as a sum of two terms,
\begin{equation}
\label{delta01}
\left\langle\overline{\delta^2(\Delta)}\right\rangle=\left\langle\overline{
\delta_0^2(\Delta)}\right\rangle+\Xi(\Delta),
\end{equation}
where the first term $\left\langle\overline{\delta_0^2(\Delta)}\right\rangle$
corresponds to the time averaged MSD (\ref{deltaUDSBMover}) obtained in the
framework of the overdamped equation (\ref{lang_sbm}) for SBM. The second term
specified in Eq.~(\ref{Xi}) accounts for the inertial effects. This term is
negative and reduces the amplitude of the time averaged MSD as compared to the
overdamped case.
For short lag times $\Delta\ll1/\gamma_0$ the ballistic regime $\left\langle
\overline{\delta^2(\Delta)}\right\rangle\simeq \Delta^2$ is obtained, as expected.
For long lag times $\Delta\gg\gamma_0^{-1}(t/\tau_0)^{1-\alpha}\gg\tau_0$ the
inertial
effects become negligible and the time averaged MSD converges to the time averaged
MSD (\ref{deltaUDSBMover}) for overdamped SBM. For superdiffusion with $\alpha>1$
and subdiffusion with values of $\alpha$ close to unity the result obtained in the
overdamped limit, Eq.~(\ref{deltaUDSBMover}), holds true for almost the entire range
of lag times $\Delta\gg\tau_0$.

\begin{figure}
\includegraphics[width=16.4cm]{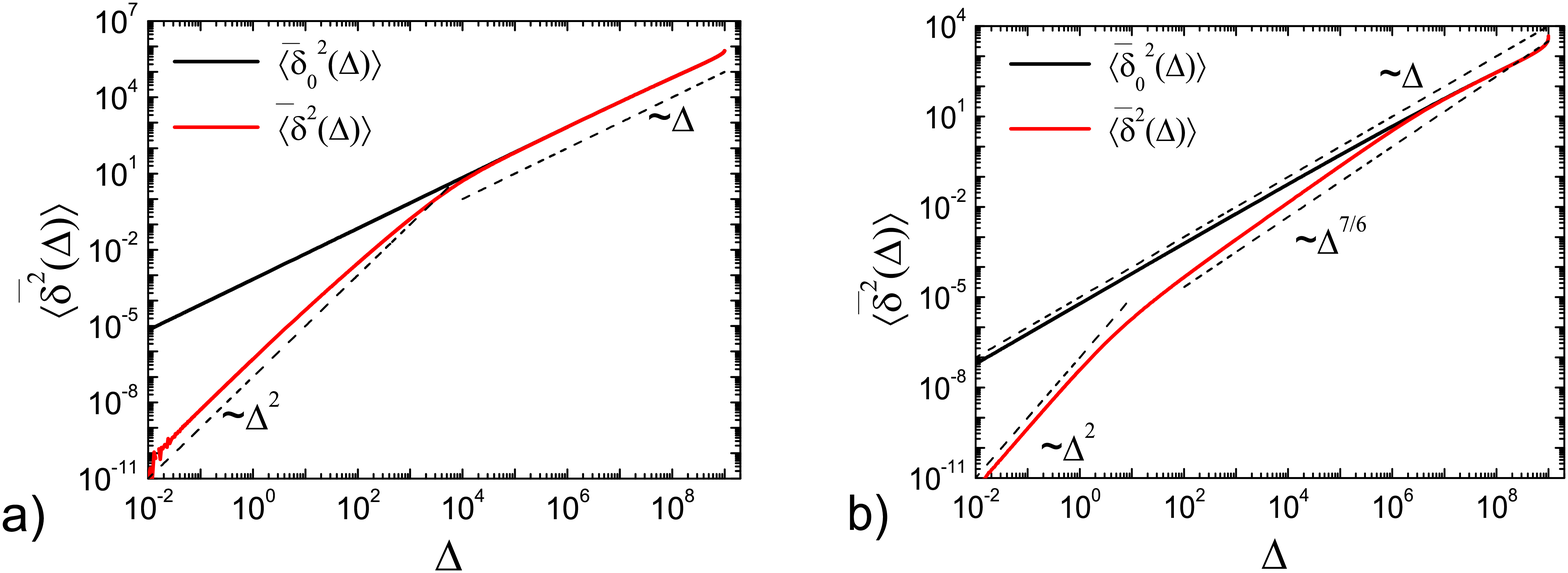}
\caption{Time averaged MSD in the overdamped limit, $\left\langle\overline{
\delta_0^2(\Delta)}\right\rangle$ from numerical integration of Eq.~(\ref{delta0})
(black line) and in the full underdamped case, $\left\langle\overline{\delta^2(
\Delta)}\right\rangle$ from Eqs.~(\ref{delta01}), (\ref{delta0}), and (\ref{Xi})
(red line). Here the trace length is $t=10^9$ and we show the cases $\alpha=1/2$
(a) and $\alpha=1/6$ (b). Dashed lines show the asymptotics at short and long
lag times. For $\alpha=1/2$ the transition between ballistic behaviour at short
times, $\left\langle\overline{\delta^2(\Delta)}\right\rangle\simeq\Delta^2$,
and the linear regime at long times, $\left\langle\overline{\delta^2(\Delta)}
\right\rangle\simeq\Delta$, is observed. For $\alpha=1/6$ an additional transient
regime becomes obvious due to long ranging effects of the underdamped motion.
The overdamped time averaged MSD is linear with
respect to $\Delta$ in both cases, $\left\langle\overline{\delta_0^2(\Delta)}
\right\rangle\simeq\Delta$. The other parameters are the same as in
Fig.~\ref{GRsupsub}. The shape of $\left\langle\overline{\delta_0^2(\Delta)}
\right\rangle$ at $\Delta\approx t$ is dominated by the pole in definition
at which $\lim_{\Delta\to t}\left\langle\overline{\delta_0^2(\Delta)}\right\rangle
=\langle x^2(t)\rangle$, see also below.}
\label{Galpha}
\end{figure}

This behaviour changes drastically for more pronounced subdiffusion. Namely, we
find that for intermediate lag times $\Delta\ll\gamma_0^{-1}(t/\tau_0)^{1-\alpha}
\ll t$ the inertial term $\Xi\left(\Delta\right)$ becomes comparable to the
overdamped term $\left\langle\overline{\delta_0^2(\Delta)}\right\rangle$, as
demonstrated in Methods. The time averaged MSD exhibits an intermediate scaling
that is not very distinctive in the case of superdiffusion, and even in the case
of subdiffusion as long as $\alpha$ is close to unity. A significant correction
occurs only for sufficiently small values of $\alpha$, that is, for pronounced
subdiffusion. This remarkable appearance of significant corrections, due to
persistent ballistic contributions, of the underdamped motion with respect to the
overdamped SBM description for subdiffusion is our first main result. It
demonstrates that in a simple yet non-stationary process the naive description of
a system in terms of the overdamped theory may lead to wrong conclusions. To our
knowledge this is the first time that such an observation for diffusive systems
is made.

In Fig.~\ref{Galpha} the results of numerical integration of Eqs.~(\ref{delta01}),
(\ref{delta0}), and (\ref{Xi}) for longer trace length $t=10^9$ are presented.
While for $\alpha=1/2$ in panel \ref{Galpha}a) the ballistic regime for $\left
\langle\overline{\delta^2(\Delta)}\right\rangle$ directly crosses over to the
asymptotic linear behaviour, for the smaller value $\alpha=1/6$ the additional
intermediate regime is distinct, Fig.~\ref{Galpha}b).
In contrast, the overdamped values of the time averaged MSD have a linear dependence
on the lag time during the whole observation time and does therefore fail to
adequately describe the behaviour of the system in the case of subdiffusion, if
only the anomalous diffusion exponent $\alpha$ is sufficiently small. We note
that the value $\alpha=1/6$ characterises the subdiffusion in a granular gas
with relative velocity dependent restitution coefficient, see section IVB. Also,
lipid molecules in a gel phase bilayer display $\alpha\approx0.16$
\cite{jae_membrane}. Small $\alpha$ values can also be tuned for the motion of
submicron beads in actin meshes \cite{wong} or for the generic motion in glassy
systems as described by the quenched trap model \cite{qel}.

\subsection*{Ultraslow underdamped scaled Brownian motion with $\alpha=0$}

We now turn to the special case of ultraslow UDSBM governed by the Langevin
equation (\ref{usbm}) with $\alpha=0$,
\begin{equation}
\frac{d^2x(t)}{dt^2}+\frac{\gamma_0}{\left(1+t/\tau_0\right)}\frac{dx(t)}{dt}=
\sqrt{\frac{2D_0}{1+t/\tau_0}}\times\frac{\gamma_0}{\left(1+t/\tau_0\right)}
\zeta(t).
\label{uusbm}
\end{equation}
In this case the velocity correlation function attains the power law time
dependence
\begin{equation}
\left\langle v(t_1)v(t_2)\right\rangle=\frac{T(0)\tau_0\gamma_0}{m(\tau_0\gamma_0
-1)}\frac{(1+t_1/\tau_0)^{\tau_0\gamma_0-2}}{(1+t_2/\tau_0)^{\tau_0\gamma_0}}.
\label{vcultraslow}
\end{equation}
The MSD can be easily calculated from this velocity correlation function, yielding
\begin{equation}
\left\langle x^2(t)\right\rangle=2D_0\tau_0\left[\log\left(1+\frac{t}{\tau_0}
\right)+\frac{1}{\tau_0\gamma_0}\left(\left(1+\frac{t}{\tau_0}\right)^{-\tau_0
\gamma_0}-1\right)\right].
\label{R2ultraslow}
\end{equation}
At times $t\ll\tau_0$ the temperature of the system does not significantly change
and the MSD behaves as if the temperature were constant, the case captured by
Eq.~(\ref{msd0}). Namely, for $t\ll 1/\gamma_0$ the MSD has the ballistic time
dependence $\left\langle x^2(t)\right\rangle=(T_0/m)t^2$ and at intermediate
times $1/\gamma_0\ll t\ll\tau_0$ normal diffusion of the form $\left\langle x^2(t)
\right\rangle=2D_0t$ is obtained. In the long time limit it scales logarithmically
as in the case of ultraslow SBM is given by Eq.~(\ref{msdsbm0}) \cite{ultraslow}.
The behaviour of the MSD in the ultraslow limit $\alpha=0$ is depicted in
Fig.~\ref{GRsupsub} by the black line.

\begin{figure}
\centerline{\includegraphics[width=12cm]{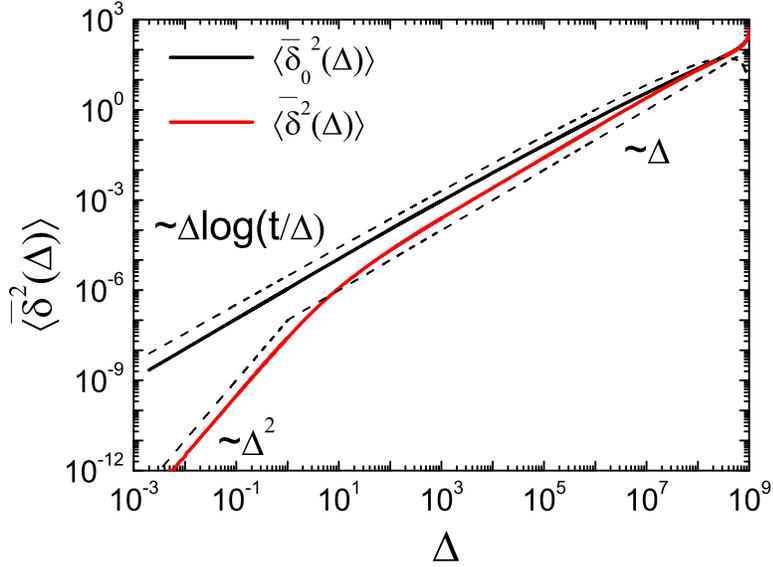}}
\caption{Time averaged MSD in the underdamped limit, $\left\langle\overline{
\delta^2(\Delta)}\right\rangle$ according to Eqs.~(\ref{delta01}),
(\ref{delta0ultraslow}), and (\ref{Xiultraslow}) (red line),
and in the overdamped limit, $\left\langle\overline{\delta_0^2(\Delta)}\right
\rangle$ according to Eq.~(\ref{delta0ultraslow}) (black line), for ultraslow
UDSBM. The measurement time is $t=10^9$, and we chose $\gamma_0=1$, $\tau_0=30$,
$D_0=1$, $m=1$, and $T_0=1$. For the underdamped time averaged MSD the crossover
between the ballistic behaviour at short times $\left\langle\overline{\delta^2(
\Delta)}\right\rangle\simeq\Delta^2$ and the linear regime at long times $\left
\langle\overline{\delta^2(\Delta)}\right\rangle\simeq\Delta$ is observed. The
overdamped time averaged MSD scales according to $\left\langle\overline{\delta_
0^2(\Delta)}\right\rangle\simeq\frac{\Delta}{t}\log\frac{t}{\Delta}$ according
to Eq.~(\ref{dusbm0}).}
\label{Gunover}
\end{figure} 

The time averaged MSD for ultraslow UDSBM may also be presented as a sum of two
terms according to Eq.~(\ref{delta01}). At short lag times $\Delta\ll1/\gamma_0$
the time averaged MSD scales ballistically, $\left\langle\overline{\delta^2(
\Delta)}\right\rangle\sim(T_0/m)\Delta^2/t$. At intermediate lag times
$\tau_0\ll\Delta\ll t/\left(\tau_0\gamma_0\right)$ the overdamped time averaged
MSD given by Eq.~(\ref{dusbm0}) is cancelled out and the underdamped time averaged
MSD has the precise linear dependence on the lag time $\Delta$
\begin{equation}
\left\langle\overline{\delta^2(\Delta)}\right\rangle\sim2D_0\tau_0\frac{\Delta}{t}.
\label{tauusbm}
\end{equation}
At longer lag times $t/\left(\gamma_0\tau_0\right)\ll\Delta\ll t$ the main term
$\delta_0(\Delta)\gg \Xi\left(\Delta\right)$ starts to dominate and the
overdamped regime according to Eq.~(\ref{dusbm0}) is observed.

This analytical result is corroborated by Fig.~\ref{Gunover} showing the comparison
between the under- and overdamped behaviours of the time averaged MSD for ultraslow
UDSBM. In the underdamped case the time averaged MSD $\left\langle\overline{\delta^2
(\Delta)}\right\rangle\simeq\Delta/t$ has the linear slope (\ref{tauusbm}) while in
the overdamped case it has the additional logarithmic correction according to
Eq.~(\ref{dusbm0}). For the parameter values used in Fig.~\ref{Gunover} the
overdamped limit is even not visible during the entire evolution of the system.
For all practical purposes, this means that the inertial corrections influence
the system's behaviour during the entire measurable time evolution. This
observation accounts for the relatively small but apparent discrepancy between
the granular gas simulations and the SBM description in Ref.~\cite{annagg}.

The persistent dominance of ballistic contributions for ultraslow UDSBM and thus
the failure of the corresponding overdamped ultraslow SBM description is our second
main result.

\section*{COMPUTER SIMULATIONS}

\begin{figure}
\includegraphics[width=16.4cm]{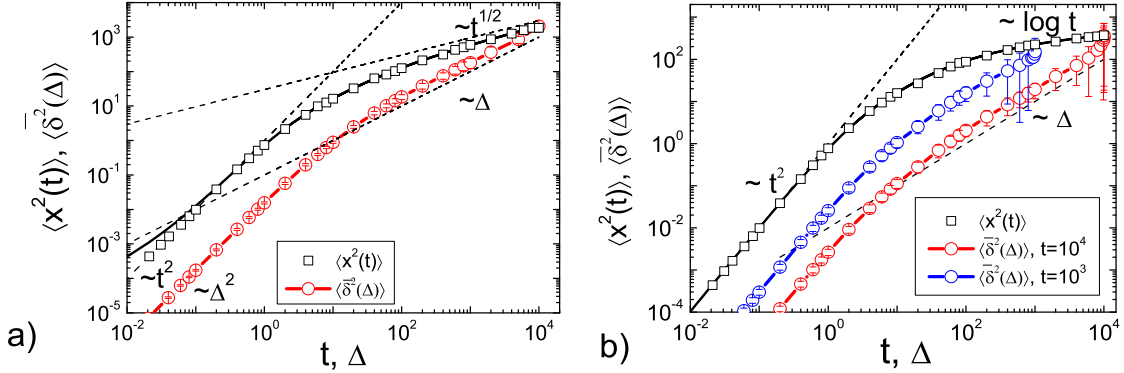}
\caption{MSD $\left\langle x^2(t)\right\rangle$ and time averaged MSD $\left
\langle\overline{\delta^2(\Delta)}\right\rangle$ obtained from computer
simulations of the corresponding finite difference analogue of the Langevin
equation for $\gamma_0=1$, $\tau_0=30$, $D_0=1$, $m=1$, $T_0=1$. We show the
cases of subdiffusion with $\alpha=1/2$ (panel a) and of ultraslow diffusion
with $\alpha=0$ (panel b). The symbols depict the simulations results of the
Langevin equations (\ref{usbm}) (a) and (\ref{uusbm}) (b).
The lines represent the analytical results (\ref{msd}) and (\ref{R2ultraslow}),
respectively.}
\label{Glang}
\end{figure}

\begin{figure}
\includegraphics[width=16.4cm]{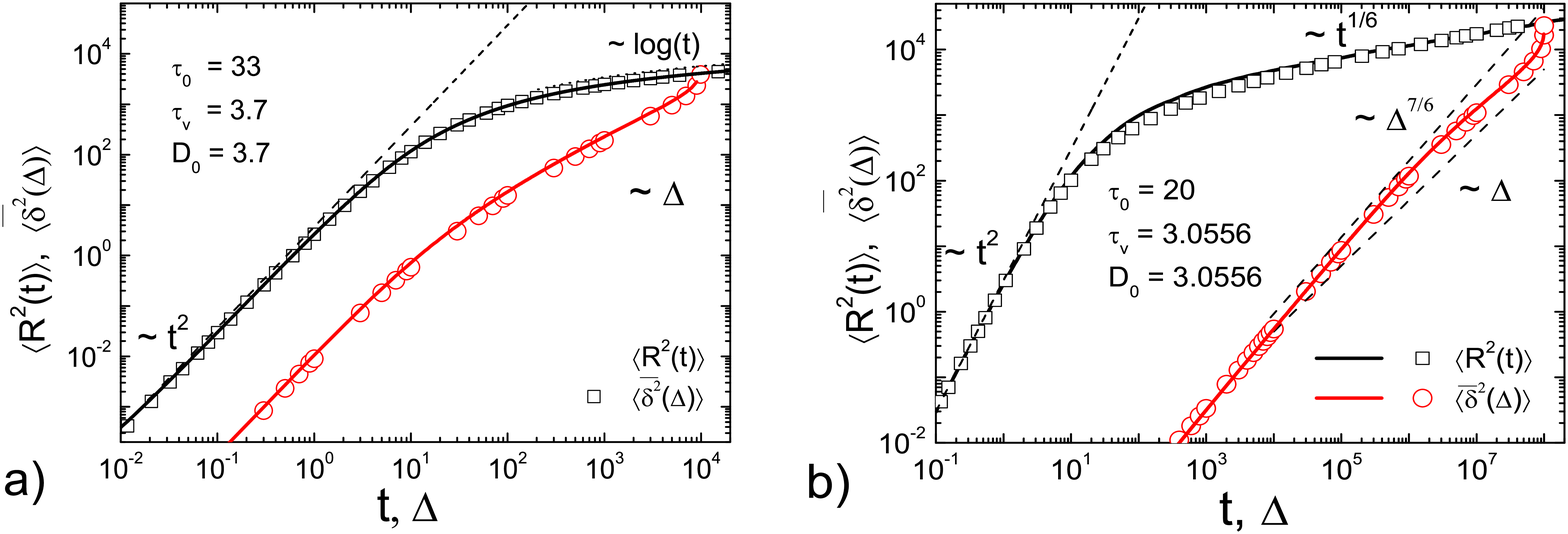}
\caption{MSD $\left\langle x^2(t)\right\rangle$ and time averaged MSD $\left
\langle\overline{\delta^2(\Delta)}\right\rangle$ from event driven computer
simulations of granular gases with constant restitution coefficient (a) and
relative velocity dependent restitution coefficient with $\alpha=1/6$ (b).
Symbols correspond
to simulation results, the lines represent the analytical results of our
UDSBM model, Eqs.~(\ref{msd}), (\ref{delta0}) and (\ref{Xi}) for panel a), and
Eqs.~(\ref{R2ultraslow}), (\ref{delta0ultraslow}), and (\ref{Xiultraslow}) for
panel b). Excellent agreement is observed.}
\label{Ggran}
\end{figure}

Here we demonstrate that our analytical results for UDSBM obtained above are indeed
confirmed by computer simulations of the corresponding finite-difference analogues
of the Langevin equations (Fig.~\ref{Glang}) and by event driven simulations of
granular gases (Fig.~\ref{Ggran}).

\subsection*{Finite difference analogue of the Langevin equation}

The finite-difference analogue of the Langevin equation may be implemented in
the following way,
\begin{subequations}
\begin{eqnarray}
v_{i+1}&=&v_i-\gamma(t_i)v_idt+\sqrt{2D(t_i)}\times\gamma_(t_i)\zeta_i
\sqrt{dt},\\
x_{i+1}&=&x_i+v_idt.
\end{eqnarray}
\end{subequations}
Here $dt=t_{i+1}-t_i$ is the time step, $v_i=v(t_i)$ and $x_i=x(t_i)$ are the
velocity and coordinate of a Brownian particle at the time $t_i$, respectively.
$\zeta_i$ is a random number distributed according to a standard normal distribution
generated using the Box-Muller transform.

The comparison of the simulations of the finite difference analogue of the
Langevin equation with theory for $\alpha=1/2$ and $\alpha=0$ are shown in
Figs.~\ref{Glang}a and \ref{Glang}b, respectively. The symbols denote the
results of the computer simulation and the lines represent the analytical
results. The simulations results are in excellent agreement with our analytical
results. At short times both MSD and time averaged MSD exhibit the expected
ballistic behaviour. At long times the MSD scales as $\langle x^2(t)\rangle\simeq
t^{1/2}$ for $\alpha=1/2$ and as $\simeq\log (t)$ for $\alpha=0$. The time
averaged MSD scales linearly at long lag times in both cases. For the
ultraslow case with $\alpha=0$ this fact underlines the remarkable and
non-negligible persistence of the ballistic effects.

\subsection*{Event driven simulations of granular gases}

In the event driven Molecular Dynamics simulations shown in Fig.~\ref{Ggran}
 we study a gas of hard sphere
granular particles of unit mass and radius, colliding respectively with constant
and viscoelastic restitution coefficients. Our simulations code is based on the
algorithm suggested in \cite{Compbook}. The particles move freely between pairwise
collisions, during the collisions the particle velocities are updated according to
certain collisional rules. The duration time of the collisions is equal to zero,
that is, the velocities of particles are updated instantaneously. Other details
of the event driven simulations are provided in \cite{annagg}. As a three
dimensional granular gas is simulated, in order to compare with our theory all
results for the moments should be divided by the factor 3.

At short times both the MSD and the time averaged MSD show a ballistic (lag) time
dependence. At long times the ensemble averaged MSD $\left\langle x^2(t)\right
\rangle$ scales according as $\simeq t^{1/6}$ for $\alpha=1/6$ and as $\simeq\log
(t)$ for $\alpha=0$ (see the two panels of Fig.~\ref{Ggran}). The time averaged
MSD $\left\langle\overline{\delta^2(\Delta)}\right\rangle$ scales linearly
for the granular gas with constant restitution coefficient, as in the case of
ultraslow UDSBM (Fig.~\ref{Ggran}a). The time averaged MSD shows a distinct
crossover behaviour for a granular gas with
velocity dependent restitution coefficient, as well as SBM with $\alpha=1/6$
(Fig.~\ref{Ggran}b). These observations demonstrate that both qualitatively
and quantitatively the behaviour of granular gases with constant and velocity
dependent restitution coefficients is fully captured by our UDSBM model. The
intermediate time deviations observed in our earlier study \cite{annagg} are thus
remedied by the inclusion of explicit long-ranging underdamped effects. The full
agreement of the UDSBM model with the granular gas dynamics is our third main
result and thus provides an interesting and easy to analytically implement model
for granular gas dynamics in the homogeneous cooling state for both constant and
velocity dependent restitution coefficients.

\section*{DISCUSSION}

We established and studied UDSBM in terms of an underdamped Langevin equation
with time dependent temperature and consequently time dependent diffusion and
damping coefficients. We derived the MSD and its time averaged analogue. As
the main findings we demonstrated that the overdamped analogue of UDSBM, the
well known SBM process, fails to adequately capture the behaviour of an UDSBM
particle even in the long time limit. Instead for pronounced subdiffusion there
exists a persistent intermediate regime for the time averaged MSD
which leads to deviations from the overdamped solution.
In the ultraslow case these corrections persist practically forever. For both
cases with $\alpha>0$ and $\alpha=0$ the corrections to the behaviour captured
by the overdamped SBM Langevin equation were corroborated by simulations of the
finite difference UDSBM Langevin equation and event driven Molecular Dynamics
simulations of cooling granular gases.
In other words, effects of inertia play a significant role even at relatively
long times and neglecting the inertial term in the Langevin equation may lead to
an incorrect description of the physical properties of the system. Given the
high accuracy achieved by modern experimental tools tracing diffusing particles
in complex environments or the possibility to run simulations over large time
windows a proper description in terms of the full underdamped dynamics is thus
highly important. This fact was demonstrated here by comparison to simulations
of granular gases with time dependent temperature (kinetic energy).

SBM can readily be extended to include an inertial term, as shown here. It can
therefore be directly compared to fractional Langevin equation motion. These two
families of anomalous stochastic processes are in some sense opposites: fractional
Langevin equation motion has stationary increments but is highly non-Markovian,
whereas UDSBM is
Markovian yet fully non-stationary. For fractional Langevin equation motion
effects of ballistic contributions were observed for the fractional
Langevin equation, leading to oscillations in the velocity correlations
\cite{eli,eli1}. Moreover, transient ageing and weak ergodicity breaking were
observed in these systems \cite{kursawe,jae_pre1,lene1}. However, these effects
decay relatively quickly. For UDSBM, in particular for small or vanishing values
of the anomalous diffusion exponent $\alpha$, these ballistic correlations turn
out to be very persistent and were shown here to be necessary to explain the full
behaviour of physical systems such as granular gases. How generic such features
are for other non-stationary anomalous diffusion processes such as heterogeneous
diffusion processes with position dependent diffusion coefficient or continuous
time random walks will therefore be an important question.

Our results demonstrate that good care is needed for the physically correct
description of anomalous diffusion processes: the naive assumption of the
equivalence of the long time behaviour and the overdamped description is not
always correct and may lead to false conclusions.

\section*{METHODS}

\section*{UNDERDAMPED SCALED BROWNIAN MOTION WITH $\alpha>0$}

The solution of the Langevin equation (\ref{usbm}) has the form
\begin{subequations}
\begin{eqnarray}
v(t)&=&v_0(t)\exp\left[-\frac{\tau_0\gamma_0}{\alpha}\left[\left(1+\frac{t}{
\tau_0}\right)^{\alpha}-1\right]\right]\\
v_0(t)&=&v(0)+\int_0^t dt^{\prime}f(t')\exp\left[\frac{\tau_0\gamma
_0}{\alpha}\left[\left(1+\frac{t^{\prime}}{\tau_0}\right)^{\alpha}-1\right]\right].
\end{eqnarray}
\end{subequations}
Here $f(t)=m\sqrt{2D(t)}\gamma(t)\zeta(t)$, the right hand side of
Eq.~(\ref{usbm}).
The velocity correlation function then yields as ($t_2>t_1$)
\begin{eqnarray}
\nonumber
\left\langle v(t_1)v(t_2)\right\rangle&=&\langle v_0^2(0)\rangle
\exp\left[-\frac{\nu}{\alpha}
\left[\left(1+\frac{t_1}{\tau_0}\right)^{\alpha}+\left(1+\frac{t_2}{\tau_0}
\right)^{\alpha}-2\right]\right]\\
\nonumber
&&+2D_0\gamma_0^2\exp\left[-\frac{\nu}{\alpha}\left[\left(1+\frac{t_1}{\tau_0}
\right)^{\alpha}-1\right]\right]\exp\left[-\frac{\nu}{\alpha}\left[\left(1+\frac{
t_2}{\tau_0}\right)^{\alpha}-1\right]\right]\\
&&\times\int_0^{t_1}dt^{\prime}\exp\left[\frac{2\nu}{\alpha}\left[\left(1+\frac{
t^{\prime}}{\tau_0}\right)^{\alpha}-1\right]\right]\left(1+\frac{t^{\prime}}{
\tau_0}\right)^{-3\left(1-\alpha\right)}.
\label{velcorr}
\end{eqnarray}
Here $\nu=\tau_0\gamma_0$. Changing the variables in the integral,
\begin{eqnarray}
\nonumber
\int_0^{t_1}dt^{\prime}\exp\left[\frac{2\nu}{\alpha}\left[\left(1+\frac{t^{\prime}
}{\tau_0}\right)^{\alpha}-1\right]\right]\left(1+\frac{t^{\prime}}{\tau_0}\right)^{
-3\left(1-\alpha\right)}\\
=\tau_0\exp\left[-\frac{2\nu}{\alpha}\right]\frac{1}{\alpha}\left(\frac{\alpha}{
2\nu}\right)^{3-\frac{2}{\alpha}}\int_{\frac{2\nu}{\alpha}}^{\frac{2\nu}{\alpha}
\left(1+\frac{t_1}{\tau_0}\right)^{\alpha}}dy e^yy^{2-\frac{2}{\alpha}}.
\end{eqnarray}
Taking into account that $e^x$ is a fast growing function, we approximate the
integral in the following way,
\begin{equation}
\int_{\frac{2\nu}{\alpha}}^{\frac{2\nu}{\alpha}\left(1+\frac{t_1}{\tau_0}\right)^{
\alpha}}dy e^yy^{2-\frac{2}{\alpha}}\sim\left(\frac{2\nu}{\alpha}\right)^{\frac{2
\alpha-2}{\alpha}}\left[\left(1+\frac{t_1}{\tau_0}\right)^{2\alpha-2}\exp\left[
\frac{2\nu}{\alpha}\left(1+\frac{t_1}{\tau_0}\right)^{\alpha}\right]-\exp\left(
\frac{2\nu}{\alpha}\right)\right].
\end{equation}
From the ensuing velocity correlation function with $\langle v_0^2(0)\rangle=
T_0/m$ we arrive at Eq.~(\ref{velcor}).

The time averaged MSD is defined as
\begin{eqnarray}
\nonumber
\left\langle\overline{\delta^2(\Delta)}\right\rangle&=&\frac{1}{t-\Delta}\int_0^{
t-\Delta}\left\langle\left[x(t^{\prime}+\Delta)-x(t^{\prime})\right]^2\right\rangle
dt'\\
&=&\frac{1}{t-\Delta}\int_0^{t-\Delta}dt^{\prime}\left[\left\langle x^2(t^{\prime}
+\Delta)\right\rangle-\left\langle x^2(t^{\prime})\right\rangle-2A\left(t^{\prime}
,\Delta\right)\right],
\label{taMSD}
\end{eqnarray}
where
\begin{eqnarray}
\nonumber
A\left(t,\Delta\right)&=&\int_0^t dt_1 \int_t^{t+\Delta}dt_2\langle v(t_1)v(t_2)
\rangle\\
\nonumber
&=&\frac{D_0}{\gamma_0} \times\left[1-\exp\left(-\frac{\nu}{\alpha}\left[\left(1+
\frac{t}{\tau_0}\right)^{\alpha}-1\right]\right)\right.\\
&&-\exp\left(\frac{\nu}{\alpha}\left[\left(1+\frac{t}{\tau_0}\right)^{\alpha}
-1\right]\right)\exp\left(-\frac{\nu}{\alpha}\left[\left(1+\frac{t+\Delta}{
\tau_0}\right)^{\alpha}-1\right]\right)\\
&&\left.+\exp\left(-\frac{\nu}{\alpha}\left[\left(1+\frac{t+\Delta}{\tau_0}
\right)^{\alpha}-1\right]\right)\right].
\nonumber
\end{eqnarray}
The integrand in Eq.~(\ref{taMSD}) attains with the velocity correlation function
(\ref{velcorr}) the following form
\begin{eqnarray}
\nonumber
\left\langle x^2(t^{\prime}+\Delta)\right\rangle-\left\langle x^2(t^{\prime})
\right\rangle-2A\left(t^{\prime},\Delta\right)&=&\frac{2D_0\tau_0}{\alpha}\left[
\left(1+\frac{t+\Delta}{\tau_0}\right)^{\alpha}-\left(1+\frac{t}{\tau_0}\right)^{
\alpha}\right]\\
&&\hspace*{-6cm}
+\frac{2D_0}{\gamma_0}\left[\exp\left(\frac{\nu}{\alpha}\left[\left(1+\frac{t}{
\tau_0}\right)^{\alpha}-1\right]\right)\exp\left(-\frac{\nu}{\alpha}\left[\left(1
+\frac{t+\Delta}{\tau_0}\right)^{\alpha}-1\right]\right)-1\right].
\end{eqnarray}
The time averaged MSD may be presented as the sum of two terms according to
Eq.~(\ref{delta01}). The first term corresponds to the time averaged MSD in the
overdamped (SBM) limit,
\begin{eqnarray}
\nonumber
\left\langle\overline{\delta_0^2(\Delta)}\right\rangle&=&\frac{2D_0\tau_0}{\alpha
\left(t-\Delta\right)}\int_0^{t-\Delta}dt^{\prime}\left[\left(1+\frac{t^{\prime}+
\Delta}{\tau_0}\right)^{\alpha}-\left(1+\frac{t^{\prime}}{\tau_0}\right)^{\alpha}
\right]\\
&&\hspace{-2cm}
=\frac{2D_0\tau_0^2}{\alpha\left(\alpha+1\right)\left(t-\Delta\right)}
\left[1+\left(1+\frac{t}{\tau_0}\right)^{\alpha+1}-\left(1+\frac{\Delta}{
\tau_0}\right)^{\alpha+1}-\left(1+\frac{t-\Delta}{\tau_0}\right)^{\alpha+1}\right].
\label{delta0}
\end{eqnarray}
The second part in the time averaged MSD reads
\begin{eqnarray}
\label{Xi}
\Xi(\Delta)=\frac{2D_0}{\gamma_0}\frac{1}{t-\Delta}\int_{0}^{t-\Delta}dt^{\prime}
\left[\exp\left(-\frac{\nu}{\alpha}\left[\left(1+\frac{t^{\prime}+\Delta}{\tau_0}
\right)^{\alpha}-\left(1+\frac{t^{\prime}}{\tau_0}\right)^{\alpha}\right]\right)
-1\right].
\end{eqnarray}

\subsection*{Short lag times: $\gamma_0,\Delta \ll t \ll \tau_0$}

From Eqs.~(\ref{delta0}) and (\ref{Xi}) we find that $\left\langle\overline{
\delta_0^2(\Delta)}\right\rangle\simeq 2D_0\Delta$ and $\Xi(\Delta)\simeq2D_0(
e^{-\gamma_0\Delta}-1)/\gamma_0$, their combination yielding for total time
averaged MSD
\begin{equation}
\left\langle\overline{\delta^2(\Delta)}\right\rangle\simeq\left<x^2(\Delta)\right>
\simeq2D_0\Delta-2D_0(1-e^{-\gamma_0\Delta})/\gamma_0
\label{final-delta2}
\end{equation}

\subsection*{Long lag times: $\tau_0 \ll \Delta \ll t$}

For the $\left\langle\overline{\delta_0^2(\Delta)}\right\rangle$ term we obtain
Eq.~(\ref{deltaUDSBMover}) from the main text, namely
\begin{equation}
\left\langle\overline{\delta_0^2(\Delta)}\right\rangle\simeq\frac{2D_0\Delta}{
\alpha \left(t/\tau_0\right)^{1-\alpha}}.
\label{deltaUSBMover-appendix}
\end{equation}
Let us now consider the additional contribution coming from Eq.~(\ref{Xi}). We
can rewrite this equation via change of variables,
\begin{eqnarray}
\Xi(\Delta)=-\frac{2D_0}{\gamma_0}\left\{1-\frac{\Delta}{t-\Delta}J\left(\lambda,
\frac{t}{\Delta}\right)\right\},
\label{Xi-append-1}
\end{eqnarray}
where the functions are defined as
\begin{equation}
J\left(\lambda,\frac{t}{\Delta}\right)=\int_{\tau_0/\Delta}^{t/\Delta-1+\tau_0/
\Delta} dx\exp[{\lambda S(x)}],
\label{J-function}
\end{equation}
as well as
\begin{equation}
S(x)=x^\alpha-(x+1)^\alpha
\label{S-function}
\end{equation}
and
\begin{equation}
\lambda=\lambda(\Delta)=\frac{\nu}{\alpha}\left(\frac{\Delta}{\tau_0}\right)^{
\alpha} \gg 1.
\label{lambda-function}
\end{equation}
We have to consider superdiffusive and subdiffusive situations separately. 

\subsection*{Superdiffusion, $\alpha>1$}

For superdiffusion the maximum of $S(x)$ is achieved at the lower limit of the
integral (\ref{J-function}), namely
\begin{equation}
\max\{S(x)\}=S(\tau_0/\Delta)\simeq-1, \quad \tau_0/\Delta\leq x\leq
t/\Delta-1+\tau_0/\Delta.
\label{maxS-equation}
\end{equation}
We estimate the integral (\ref{J-function}) with the method of steepest descent,
\begin{equation}
J\simeq-\frac{\exp[{\lambda S(\tau_0/\Delta)}]}{\lambda S'(\tau_0/\Delta)},
\,\,\, S(\tau_0/\Delta)\simeq-1, \,\,\, S'(\tau_0/\Delta)\simeq-\alpha.
\label{eq-J-steep}
\end{equation}
Therefore, we find that $J$ gives an exponentially small contribution to $\Xi(
\Delta)$ and
\begin{equation}
\Xi(\Delta)=-2D_0/\gamma_0,
\label{Xi-2-append}
\end{equation}
that is
\begin{equation}
\frac{|\Xi(\Delta)|}{\left\langle\overline{\delta_0^2(\Delta)}\right\rangle}
\simeq\frac{\alpha}{\gamma_0 \Delta} \left(\frac{\tau_0}{t}\right)^{\alpha-1}\ll1.
\label{eq-append-ratio}
\end{equation}
Thus, the overdamped result for the time averaged MSD provides the correct result
in the superdiffusive case.

\subsection*{Subdiffusion, $0<\alpha<1$}

In the subdiffusive case the maximum of $S(x)$ is achieved at the upper limit of
the integral (\ref{J-function}),
\begin{equation}
\max\{S(x)\}\simeq S(t/\Delta-1)\simeq-\frac{\alpha\Delta^{1-\alpha}}{(t-\Delta)^{
1-\alpha}}.
\label{maxS-equation-2}
\end{equation}
For longer lag times, such that
\begin{equation}
\lambda S(t/\Delta-1)\simeq\gamma_0\Delta(\tau_0/t)^{1-\alpha}\gg1,
\label{S-condition}
\end{equation}
that is $\gamma_0^{-1}(t/\tau_0)^{1-\alpha}\ll\Delta\ll t$, the contribution of
$J$ is again exponentially small and we have---similarly to the superdiffusive
case---that
\begin{equation}
\Xi(\Delta)=-2D_0/\gamma_0.
\label{Xi-2-append-super}
\end{equation}
Thus, due to Eq.~(\ref{S-condition}) we see that
\begin{equation}
\frac{|\Xi(\Delta)|}{\left\langle\overline{\delta_0^2(\Delta)}\right\rangle}\sim
\frac{\alpha}{\gamma_0\Delta}\left(\frac{t}{\tau_0}\right)^{1-\alpha} \ll1,
\label{eq-append-ratio-super}
\end{equation}
and the time averaged MSD corresponds to the overdamped approximation. In
contrast, for shorter lag times,
\begin{equation}
\tau_0\ll\Delta\ll\frac{1}{\gamma_0}\left(\frac{t}{\tau_0}\right)^{1-\alpha},
\label{shorter-lags-super}
\end{equation}
the method of the steepest descent is not valid. We may roughly estimate the lower
bound of $|\Xi(\Delta)|$ as
\begin{equation}
|\Xi(\Delta)|_{\min}=\frac{2D_0}{\gamma_0}\left\{1-\frac{\Delta}{t-\Delta}\int_0^{
t/\Delta-1}dx\exp[\lambda\cdot\max\{S(x)\}]\right\}=\frac{2D_0}{\gamma_0}\lambda
S\left(\frac{t}{\Delta}-1\right),
\label{lower-Xi-super}
\end{equation}
and thus
\begin{equation}
\frac{|\Xi(\Delta)|_{\min}}{\left<\overline{\delta_0^2(\Delta)}\right>}
=\alpha.
\end{equation}
This estimate shows that in the domain of variables (\ref{shorter-lags-super}) the
contributions to the time averaged MSD stemming from the terms $\left\langle
\overline{\delta_0^2(\Delta)}\right\rangle$ and $\Xi(\Delta)$ are of comparable
magnitude, and thus inertial effects cannot be neglected in the consideration.  

\section*{ULTRASLOW UNDERDAMPED SCALED BROWNIAN MOTION WITH $\alpha=0$}

Ultraslow UDSBM corresponds to the case $\alpha=0$ in which the velocity
correlation function (\ref{vcultraslow}) and the MSD (\ref{R2ultraslow}) may be
obtained from the results of the previous section in the limit $\alpha\to0$ taking
into account that $\lim_{\alpha\to0}\frac{c^{\alpha}-1}{\alpha}=\log c$.

The first term of the time averaged MSD corresponds to the time averaged MSD for
ultraslow SBM,
\begin{eqnarray}
\nonumber
\left<\overline{\delta_0^2(\Delta)}\right>&=&\frac{2D_0\tau_0}{t-\Delta}
\int_0^{t-\Delta}dt^{\prime}\left[\log\left(1+\frac{\Delta}{\tau_0}+\frac{t'}{\tau
_0}\right)-\log\left(1+\frac{t'}{\tau_0}\right)\right]\\
\nonumber
&=&\frac{2D_0\tau_0^2}{t-\Delta}\left\{\left(1+\frac{t}{\tau_0}\right)\log\left(
1+\frac{t}{\tau_0}\right)\right.\\
&&-\left.\left(1+\frac{\Delta}{\tau_0}\right)\log\left(1+\frac{\Delta}{\tau_0}
\right)-\left(1+\frac{t-\Delta}{\tau_0}\right)\log\left(1+\frac{t-\Delta}{\tau_0}
\right)\right\}.
\label{delta0ultraslow}
\end{eqnarray}
The second term may be derived analogously to the previous section,
\begin{eqnarray}
\Xi(\Delta)=\frac{2D_0}{\gamma_0(t-\Delta)}\int_0^{t-\Delta}dt^{\prime}\left[
\frac{\left(1+\frac{t'}{\tau_0}
\right)^\nu}{\left(1+\frac{t'+\Delta}{\tau_0}
\right)^\nu}-1\right]<0,
\label{Xiultraslow}
\end{eqnarray}
where we took into account that $\nu=\tau_0\gamma_0\gg1$. In what follows we
consider separately the limits of short and long lag times.  

\subsection*{Short lag times, $\Delta \ll t \ll \tau_0$} 

From Eqs.~(\ref{delta0ultraslow}) and (\ref{Xiultraslow}) we find
\begin{equation}
\left\langle\overline{\delta_0^2(\Delta)}\right\rangle\simeq 2D_0\Delta,
\label{delta02}
\end{equation} 
and
\begin{equation}
\Xi(\Delta)\simeq-\frac{2D_0}{\gamma_0}\left[1-e^{-\gamma_0\Delta}\right].
\label{Theta2}
\end{equation}
By combining expressions (\ref{delta02}) and (\ref{Theta2}) we get the time
averaged MSD
\begin{equation}
\left\langle\overline{\delta^2(\Delta)}\right\rangle\simeq\left<x^2(\Delta)
\right>\simeq 2D_0\Delta-\frac{2D_0}{\gamma_0}\left[1-e^{-\gamma_0\Delta}\right],
\label{equival-x2-d2}
\end{equation}
as expected for short lag times, see Eq.~(\ref{ddd0}) of the main text. Note the
approximate sign in Eq.~(\ref{equival-x2-d2}) because it is valid up to terms
that are smaller by the factor $t/\tau_0$. 

\subsection*{Long lag times, $\tau_0 \ll \Delta \ll t$} 

The contribution given by relation (\ref{delta0ultraslow}) can be calculated
directly,
\begin{equation}
\left\langle\overline{\delta_0^2(\Delta)}\right\rangle\simeq2D_0\tau_0\frac{
\Delta}{t}  \left(1+\log\frac{t}{\Delta}\right).
\label{d22}
\end{equation}
Changing variables in the integrand of Eq.~(\ref{Xiultraslow}) we rewrite
it as
\begin{eqnarray}
\Xi\left(\Delta\right)\simeq-\frac{2D_0}{\gamma_{0}}\left[1-\frac{\Delta}{
t-\Delta}I\left(t,\Delta\right)\right], 
\label{T22}
\end{eqnarray}
where we define
\begin{equation}
I\left(t,\Delta\right)=\int_{\frac{\Delta}{t-\Delta+\tau_0}}^{\frac{\Delta}{
\tau_0}}\frac{dy}{y^2\left(1+y\right)^{\nu}}.
\label{intI}
\end{equation}
Since the integrand is decaying fast at $y\to\infty$, we can safely replace the
upper limit of the integral by $\infty$. Moreover we can neglect the term $\tau_
0$ at the lower integration limit. Then we integrate by parts twice in order to
extract the main terms such that
\begin{eqnarray}
\nonumber
I\left(t,\Delta\right)&\approx&\int_{\frac{\Delta}{t-\Delta}}^{\infty}\frac{dy}{
y^2\left(1+y\right)^{\nu}}\\
\nonumber
&=&-\left.\frac{1}{y\left(1+y\right)^{\nu}}\right|_{\frac{
\Delta}{t-\Delta}}^{\infty}-\nu\int_{\frac{\Delta}{t-\Delta}}^{\infty}
\frac{dy}{y\left(1+y\right)^{\nu+1}}\\
\nonumber
&=&\frac{t-\Delta}{\Delta}\frac{1}{\left(1+\frac{\Delta}{t-\Delta}\right)^\nu}
+\nu\left(1+\frac{\Delta}{t-\Delta}\right)^{-\nu-1}\log\left(\frac{\Delta}{t-
\Delta}\right)\\
&&-\nu(\nu+1)\int_{\frac{\Delta}{t-\Delta}}^{\infty}\frac{\log(y)dy}{\left(1+y
\right)^{\nu+2}}
\label{intIapprox}
\end{eqnarray}
The integrand in the last term of the right hand side has an integrable divergence
at zero, thus we can safely put the lower limit to zero and use \cite{prudnikov}
\begin{equation}
\int_{0}^{\infty} dy\frac{\log y}{\left(1+y\right)^{\nu+2}}=-\frac{1}{\nu+1}
\left[\gamma+\frac{1}{\nu}+\psi\left(\nu\right)\right],
\label{psi-formula}
\end{equation}
where $\gamma=0.5772\ldots$ is Euler's constant and $\psi\left(z\right)=\frac{d
\log[\Gamma(z)]}{dz}$ is the digamma function. After plugging (\ref{psi-formula})
into (\ref{intIapprox}) and then (\ref{intIapprox}) into (\ref{T22}) we get
\begin{eqnarray}
\nonumber
\Xi\left(\Delta\right)&=&-\frac{2D_0}{\gamma_{0}}\left[1-\left(1-\frac{\Delta}{t}
\right)^\nu-\frac{\nu\Delta}{t}\left(1-\frac{\Delta}{t}\right)^\nu\log\left(\frac{
\Delta}{t-\Delta}\right)\right.\\
&&\left.-\left(\frac{\Delta}{t-\Delta}\right)\nu C(\nu)\right],
\label{Xifinal2}
\end{eqnarray}
where the following definition is introduced 
\begin{equation}
C(\nu)=\gamma+\frac{1}{\nu}+\psi\left(\nu\right).
\label{c-nu}
\end{equation}
Equation (\ref{Xifinal2}) exhibits two different behaviours in the long time
limit considered here. Thus, for $\tau_0\ll\Delta\ll t/\nu$ we find
\begin{equation}
\Xi\left(\Delta\right)\simeq-2D_0\tau_0\frac{\Delta}{t}\left[1+\log\left(\frac{t}{
\Delta}\right)-C(\nu)\right],
\label{TTT1}
\end{equation}
and by combining (\ref{d22}) and (\ref{TTT1}) we observe the cancellation of the
main terms in  $\left\langle\overline{\delta_0^2(\Delta)}\right\rangle$ and
$\Xi\left(\Delta\right)$, resulting for the time averaged MSD in
\begin{equation}
\left\langle\overline{\delta^2(\Delta)}\right\rangle\approx2D_0\tau_0 C(\beta)
\frac{\Delta}{t}.
\label{tamsd-final-eq}
\end{equation}
For longer lag times $\tau_0 \ll t/\nu \ll \Delta \ll t$ Eq. (\ref{Xifinal2})
yields
\begin{equation}
\Xi\left(\Delta\right)\simeq2D_0\tau_0C(\beta)\frac{\Delta}{t}\ll\left\langle
\overline{\delta_0^2(\Delta)}\right\rangle \simeq2D_0\tau_0\frac{\Delta}{t}
\log\left(\frac{t}{\Delta}\right).
\label{TTT2}
\end{equation}
Thus, in this case the main term of $\left\langle\overline{\delta_0^2(\Delta)}
\right\rangle$ is not cancelled out, and the overdamped regime (\ref{msdsbm0})
of the main text is observed.

\section*{Acknowledgements}

The authors thank N. V. Brilliantov for stimulating discussions. The
simulations were run at the Chebyshev supercomputer of Moscow State
University.

\section*{Author contributions}

ASB, AVC, AGC, HS, IMS, and RM conceived and carried out the research,
ASB, AVC, AGC, HS, IMS, and RM wrote and reviewed the paper.

\section*{Competing financial interests}

The authors declare no competing financial interests.

\end{document}